\documentclass[
seceq,
preprint]{ptptex}

\newcommand{\Dslash}{{\not}\kern-0.05em D}
\newcommand{\dslash}{{\not}\kern+0.1em\partial}
\newcommand{\tr}{\mathop{\rm tr}\nolimits}

\newcommand{\Pf}{\mathop{\rm Pf}\nolimits}
\newcommand{\ind}{\mathop{\rm index}\nolimits}

\newcommand{\SO}{\mathop{\rm SO}}
\newcommand{\U}{\mathop{\rm {}U}}

\newcommand{\rmd}{{\rm d}}

\preprintnumber[3cm]{
RIKEN-TH-62
}

\title{
Gauge Anomaly associated with the Majorana Fermion in $8k+1$ dimensions%
}

\author{
Masashi \textsc{Hayakawa}\footnote{E-mail: haya@riken.jp}
and Hiroshi \textsc{Suzuki}\footnote{E-mail: hsuzuki@riken.jp}
}

\inst{
Theoretical Physics Laboratory, RIKEN, Wako 2-1, Saitama 351-0198, Japan
}

\abst{
Using an elementary method, we show that an odd number of Majorana fermions in
$8k+1$ dimensions suffer from a gauge anomaly that is analogous to the
Witten global gauge anomaly. This anomaly cannot be removed without
sacrificing the perturbative gauge invariance. Our construction of
higher-dimensional examples ($k\geq1$) makes use of the $\SO(8)$ instanton
on $S^8$.
}

\begin{document}

\maketitle

\section{Introduction}
The Majorana fermion in $8k+1$-dimensional Minkowski spacetime is very
peculiar. The charge conjugation matrix is symmetric $C^T=+C$ in $8k+1$
dimensions and only possible Lorentz invariant mass term for a single
Majorana fermion, $m\psi^TC^{-1}\psi$, identically vanishes. This implies
that one cannot apply the gauge invariant Pauli-Villars regularization to a
single Majorana fermion. This situation generally persists for odd number of
Majorana fermions.

A similar phenomenon is also found with the lattice
regularization~\cite{Inagaki:2004ar}. The most impressive example is given by
the 1-dimensional ($k=0$) case, i.e., quantum mechanics, for which a lattice
action of a free Majorana fermion~$\widetilde\psi$ in the momentum space
would be
\begin{equation}
   \int_{-\pi/a}^{\pi/a}\rmd p\,\widetilde\psi^T(-p)
   \widetilde D(p)\widetilde\psi(p),
\label{one}
\end{equation}
where $a$ denotes the lattice spacing. The variable~$\widetilde\psi$ has a
gauge index but no spinor index. Since $\widetilde\psi$ is Grassmann-odd,
the kernel $\widetilde D(p)$ must be odd under $p\leftrightarrow-p$. Also,
$\widetilde D(p)$ must be periodic in the Brillouin zone
$\widetilde D(p+2\pi/a)=\widetilde D(p)$ for locality. These two
requirements, however, imply that the kernel inevitably possesses a zero at
$p=\pi/a$ which corresponds to the species doubler. Thus one cannot write
down a lattice action of a free single Majorana fermion which is consistent
with locality, at least in the form of the ansatz~(\ref{one}).

These ``phenomenological'' observations strongly suggest the existence of
some sort of anomaly associated to the Majorana fermion in
$8k+1$~dimensions. Since there is no perturbative (or local) gauge anomaly
in these {\it odd\/} dimensions, one expects that something similar to the
global gauge anomaly in even-dimensional
spaces~\cite{Witten:fp,Elitzur:1984kr} occurs. This possibility has been
suggested in ref.~\citen{Alvarez-Gaume:1983ig} and the global
{\it gravitational\/} anomaly~\cite{Witten:1985xe} for the Majorana fermion
in these dimensions has been demonstrated. Also the global {\it gauge\/}
anomaly associated to the Majorana fermion in 1~dimension ($k=0$) has been
known~\cite{Friedan:1983xr}. For example, supersymmetric quantum mechanics
which describes a spinning particle in curved spacetime suffers from the
global ``gauge'' anomaly, when the spacetime does not admit a spin structure.
To our knowledge, however, any examples of the global gauge anomaly
associated to the Majorana fermion in higher ($k\geq1$) dimensions have not
been given.

In this paper, we provide an example of the gauge anomaly for the Majorana
fermion in $8k+1$~dimensions in the
$(\mathbf{N},\mathbf{8_v},\ldots,\mathbf{8_v})$ representation of the gauge
group $\SO(N)\times\SO(8)^k$ ($N\geq3$). After starting with generalities
concerning with the problem, we describe our strategy to demonstrate the
existence of gauge anomaly in that system with base space of topology
$S^1\times S^8\times\cdots\times S^8$. 

The Euclidean action for the Majorana fermion is given by
\begin{equation}
   S=\int\rmd^{8k+1}x\,\psi^T(x)\Dslash_{8k+1}\psi(x),
\end{equation}
where the covariant derivative is defined by
$\Dslash_{8k+1}=\sum_{\mu=1}^{8k+1}\gamma_\mu\{\partial_\mu+A_\mu(x)\}$ and
we have chosen a representation in which all $\gamma_\mu$ are real and
symmetric. The gauge potential $A_\mu(x)$ takes the value in the fundamental
representation of the Lie algebra of~$\SO(N)\times\SO(8)^k$ and
$A_\mu^*=A_\mu=-A_\mu$. The Dirac operator $\Dslash_{8k+1}$ is thus real and
anti-symmetric. The partition function is formally given by the Pfaffian of
the Dirac operator,
\begin{equation}
   \Pf\{i\Dslash_{8k+1}\},
\label{three}
\end{equation}
whose square is the Dirac determinant
\begin{equation}
   \det\{i\Dslash_{8k+1}\}=(\Pf\{i\Dslash_{8k+1}\})^2.
\label{four}
\end{equation}
The Dirac determinant can be defined in a gauge-invariant manner by taking an
appropriately regularized product of all eigenvalues~$\lambda_n$ of
$i\Dslash_{8k+1}$:
\begin{equation}
   i\Dslash_{8k+1}\varphi_n(x)=\lambda_n\varphi_n(x).
\label{five}
\end{equation}
We remark that all the eigenvalues~$\lambda_n$ are real and, because
$\varphi_n^*$ is an eigenfunction with the eigenvalue $-\lambda_n$, nonzero
eigenvalues come in pairs $(\lambda_n,-\lambda_n)$. Equation~(\ref{four})
indicates that the Majorana Pfaffian is defined by the product of either
the eigenvalue $\lambda_n$ or the eigenvalue $-\lambda_n$ for each~$n$. The
choice of $\lambda_n$ or $-\lambda_n$, however, leads to an ambiguity in the
sign of the product. We would like to remove this ambiguity so that the Majorana
Pfaffian is a smooth function of gauge-field configurations to ensure that the
Schwinger-Dyson equations hold~\cite{Witten:fp}. We can define
$\Pf\{i\Dslash_{8k+1}\}[A]$ for a particular gauge field~$A$ as the product of
(say) the positive $\lambda_n$. Then, by requiring the smoothness of the
Pfaffian, there is no sign ambiguity for other gauge-field configurations in
each topological sector.

Our prime interest regards the possibility that the Majorana
Pfaffian~(\ref{three}) may not be invariant under a certain gauge
transformation, defined as
\begin{equation}
   A_\mu^g(x)\equiv g(x)^{-1}\{\partial_\mu+A_\mu(x)\}g(x).
\label{six}
\end{equation}
Because $A$ and $A^g$ give an identical set of eigenvalues, the Pfaffian is
gauge invariant up to a sign:
\begin{equation}
   \Pf\{i\Dslash_{8k+1}\}[A^g]=\pm\Pf\{i\Dslash_{8k+1}\}[A].
\label{seven}
\end{equation}
By noting the smoothness of the Pfaffian, one can show that the sign in
Eq.~(\ref{seven}) does not change under any smooth deformation of the gauge
field~$A$; the sign depends only on~$g$ and, possibly, on the topology of~$A$.
We may, therefore, adopt a convenient form of the gauge field~$A$ when
computing the sign in Eq.~(\ref{seven}).

In \S~2, we first show the existence of a gauge anomaly in as $\SO(N)$
gauge system with an odd number of Majorana fermions on ${\mathbb R}^1$ or
$S^1$ ($k=0$). In \S~3, we find the gauge field $A$ and gauge
transformation~$g$ which lead to a minus sign in Eq.~(\ref{seven}) through
the mechanism in the 1-dimensional system presented in \S~2, supplemented with
the appropriate number of $\SO(8)$ instantons\cite{Grossman:1984pi}.
Through its construction, the base manifold is found to take the form of the
topology of $S^1\times S^8\times\cdots\times S^8$. A conclusion is given in
\S~4.

\section{One-dimensional case}
For $k=0$, the gamma matrix is a scalar, specifically, $\gamma_1=1$, and the
Dirac operator can thus be written as $D_1=\partial_1+A_1$. We take the
following form of the gauge potential:
\begin{equation}
   A_1(x_1)=A_1^1(x_1)T^1,\quad T^1=
   \begin{pmatrix} 0&1& &      & \\
                  -1&0& &      & \\
                    & &0&      & \\
                    & & &\ddots& \\
                    & & &      &0
   \end{pmatrix}.
\label{eight}
\end{equation}
Here $T^1$ is one of the generators of $\SO(N)$. For this special
configuration, we can explicitly solve the eigenvalue problem~(\ref{five}). [See
Exercise~5.8 of Ref.~\citen{Jackiw:1983nv} for a similar problem in a $\U(1)$
theory.] We impose the periodic boundary condition
$\varphi_n(+L/2)=\varphi_n(-L/2)$ on a finite interval
$-L/2\leq x_1\leq+L/2$. The eigenvalues are then given by
\begin{equation}
   \lambda_nL=\pm a+2\pi n,\quad n\in\mathbb{Z},
\label{nine}
\end{equation}
where
\begin{equation}
   a=\int_{-L/2}^{+L/2}\rmd x_1\,A_1^1(x)
   =-{1\over2}\int_{-L/2}^{+L/2}\rmd x_1\,\tr\{T^1A_1(x_1)\},
\label{ten}
\end{equation}
and
\begin{equation}
   \lambda_nL=2\pi n,\qquad n\in\mathbb{Z}.
\label{eleven}
\end{equation}
The latter eigenvalues are $N-2$-fold degenerate.

The Dirac determinant is given by the product of all the eigenvalues in
Eqs.~(\ref{nine}) and~(\ref{eleven}). This product can be regularized by
dividing it by the determinant of the free Dirac operator:
\begin{equation}
   {\det'\{iD_1\}[A]\over\det'\{iD_1\}[0]}
   =-{a^2\over L^2}\prod_{n=1}^\infty\left(1-{a^2\over4\pi^2n^2}\right)^2
   =-{4\over L^2}\sin^2\left({a\over2}\right),
\label{twelve}
\end{equation}
where the prime indicates that zero eigenvalues independent of the gauge field
are omitted from the product. From this, we have
\begin{equation}
   {\Pf'\{iD_1\}[A]\over\Pf'\{iD_1\}[0]}
   ={2i\over L}\sin\left({a\over2}\right),
\label{thirteen}
\end{equation}
because this defines a smooth function of the gauge field~$A$. If the
Pfaffian were defined by the absolute value $(2i/L)|\sin(a/2)|$ instead, it
would develop a cusp at~$a=0$, and, consequently, the Schwinger-Dyson
equations would not hold.

Now, we consider a particular gauge transformation,
\begin{equation}
   g(x_1)=e^{\theta^1(x_1)T^1}.
\label{fourteen}
\end{equation}
To conclude the existence of a global anomaly, it is sufficient to show
that the sign in Eq.~(\ref{seven}) is minus for the gauge transformation
which satisfies the boundary condition $g(-L/2)=g(+L/2)=1$. The real
function~$\theta^1(x_1)$ in Eq.~(\ref{fourteen}) thus satisfies
\begin{equation}
   \theta^1(-L/2)=2\pi w_-,\quad\theta^1(+L/2)=2\pi w_+,\quad
   w_\pm\in\mathbb{Z}.
\label{fifteen}
\end{equation}
From Eqs.~(\ref{six}), (\ref{fourteen}) and~(\ref{fifteen}), it is found that
the integral~(\ref{ten}) changes under the gauge transformation as
\begin{equation}
   a^g=-{1\over2}\int_{-L/2}^{+L/2}\rmd x_1\,\tr\{T^1A_1^g(x_1)\}
   =a+2\pi w,
\label{sixteen}
\end{equation}
where $w\equiv w_+-w_-$. From Eq.~(\ref{thirteen}), we see that the Pfaffian
actually changes sign under the gauge transformation
\begin{equation}
   \Pf\{iD_1\}[A^g]=(-1)^w\Pf\{iD_1\}[A]
\label{seventeen}
\end{equation}
when the ``winding number'' $w$ of the gauge transformation is odd. As we
noted in the introduction, this relation should hold for all gauge-field
configurations~$A$ that can be smoothly deformed into the above particular
configuration~(\ref{eight}).

It is legitimate to regard $(-1)^w$ in Eq.~(\ref{seventeen}) as a gauge
anomaly, because it cannot be removed by any local counterterm. More
precisely, it can be removed {\it only by sacrificing the local (or
perturbative) gauge invariance}. In fact, the anomaly~$(-1)^w$ is removed by
adding the local term
\begin{equation}
   \Delta S={i\over4}\int_{-L/2}^{+L/2}\rmd x_1\,\tr\{T^1 A_1(x_1)\}
\end{equation}
to the action, because this term changes as $\Delta S\to\Delta S-i\pi w$
under the gauge transformation~(\ref{fourteen}) [see Eq.~(\ref{sixteen})]. This
term, however, manifestly breaks the local $\SO(N)$ gauge symmetry for
$N\geq3$. In this respect, the present gauge anomaly should be clearly
distinguished from the apparent global gauge anomaly associated with the Dirac
fermion in an odd number of dimensions~\cite{Redlich:1983kn}, which can always
be removed by the Chern-Simons term. We conclude that a single Majorana fermion
in 1~dimension suffers from the global gauge anomaly and, in general, cannot be
defined in a gauge invariant way. This conclusion also holds in the case that
the number of Majorana fermions is odd.

In order to find a higher-dimensional analogue of the above-described
phenomenon, it is useful to observe the anomaly from the point of view of a
flow of eigenvalues of $iD_1$. For this purpose, introduce an additional
coordinate~$x_0$ ($-\infty<x_0<+\infty$), and the one-parameter family of gauge
fields, $A_1(x_0,x_1)$, which adiabatically interpolates between our previous
gauge field~(\ref{eight}) in the limit $x_0\to-\infty$ and the gauge
transformed one $A^g$ in the limit $x_0\to+\infty$. The Pfaffian for
$x_0\to-\infty$, $\Pf\{iD_1\}[A]$, is defined by the product of (say) positive
eigenvalues of $iD_1$. The Pfaffian for $x_0\to+\infty$, $\Pf\{iD_1\}[A^g]$, is
then uniquely given by the product of the eigenvalues of the eigenfunctions
adopted for $x_0\to-\infty$. The global gauge anomaly occurs as a result of a
flow of eigenvalues along~$x_0$ in which an odd number of positive eigenvalues
of $iD_1$ for $x_0\to-\infty$ becomes negative for $x_0\to+\infty$.
(See~Ref.~\citen{Witten:fp}.) Such a flow of eigenvalues is closely related to
the number of right-handed zero modes of the Dirac operator in 2~dimensions,
\begin{equation}
   \Dslash_2=\sigma^2{\partial\over\partial x_0}
   +\sigma^1D_1,
\label{nineteen}
\end{equation}
where $\sigma^1$ and $\sigma^2$ denote the Pauli matrices, and the gauge
potential in~$D_1$ is $A_1(x_0,x_1)$. A zero mode~$\Psi$, for which
$\Dslash_2\Psi=0$, with right-handed chirality, expressed as
$\sigma^3\Psi=+\Psi$, satisfies ${\partial\Psi\over\partial x_0}=iD_1\Psi$. In
the adiabatic approximation, this is solved as
\begin{equation}
   \Psi(x_0,x_1)=
   \begin{pmatrix}
   \exp\{+\int_0^{x_0}\rmd x_0'\,\lambda_n(x_0')\}\varphi_n(x_0,x_1)\\
   0
   \end{pmatrix}.
\end{equation}
This wave function is normalizable iff $\lambda_n(x_0)$ is positive for
$x_0\to-\infty$ and negative for $x_0\to+\infty$. Hence, a flow of eigenvalues
exhibits the global gauge anomaly iff the number of right-handed zero modes
of $\Dslash_2$ is odd. The above study of the 1-dimensional case shows that the
number of right-handed zero modes of $\Dslash_2$ is odd for~$g$ in
Eq.~(\ref{fourteen}) with an odd winding number~$w$.

Similarly, left-handed zero modes of $\Dslash_2$ correspond to eigenfunctions
for which $\lambda_n(x_0)$ is negative for $x_0\to-\infty$ and positive for
$x_0\to+\infty$. As noted above, eigenvalues of $D_1$ come in pairs,
$\pm\lambda_n$. Thus the number of right-handed zero modes of $\Dslash_2$ is
the same as the number of left-handed ones. In other words, the index of
$\Dslash_2$ identically vanishes: $\ind\{i\Dslash_2\}=0$.

\section{$8k+1$-dimensional case}
Let us now turn to the $8k+1$-dimensional problem. For this, we introduce
the Dirac operator in $8k+2$~dimensions
\begin{eqnarray}
   &&\Dslash_{8k+2}=\Dslash_2\otimes 1+\sigma^3\otimes\Dslash_{8k},
\nonumber\\
   &&\Dslash_{8k}=\sum_{\mu=2}^{8k+1}\gamma_\mu(\partial_\mu+A_\mu).
\end{eqnarray}
We choose such a representation that all $8k$-dimensional gamma matrices
$\gamma_\mu$ ($\mu=2$, 3, \dots, $8k+1$) are real. The corresponding
chiral matrices are given by
\begin{equation}
   \Gamma^{(8k+2)}=\sigma^3\otimes\gamma^{(8k)},\quad
   \gamma^{(8k)}=\gamma_2\gamma_3\cdots\gamma_{8k+1}.
\label{twentytwo}
\end{equation}

To show that the minus sign is realized in Eq.~(\ref{seven}), it is enough to
find $(A_\mu,g)$ such that the number of right-handed zero modes of
$\Dslash_{8k+2}$ is odd, as in the $k=0$ case in the previous
section.\footnote{It can be confirmed that $\Dslash_{8k+2}\Psi=0$ with
$\Gamma^{(8k+2)}\Psi=+\Psi$ is equivalent to the non-trivial flow of an
eigenvalue of~$i\Dslash_{8k+1}$, in which the last gamma matrix is given
by~$\gamma^{(8k)}$. It can also easily be shown that the number of zero modes
of $\Dslash_{8k+2}$ modulo~4 is invariant under a smooth deformation of the
gauge field and is either 0 or $2\pmod4$; this thus defines an index. What we
want to find is an appropriate case for which this index is $2\pmod4$.} For
this purpose, we choose $A_1(x_0,x_1)$ to be identical to the above
1-dimensional example with $w$ odd. It thus takes values in the Lie
algebra of~$\SO(N)$. Other components of the gauge field, $A_\mu$ ($\mu=2$, 3,
\dots, $8k+1$), are assumed to take values in the Lie algebra of
$\SO(8)^k$ and depend only on $x_2$, $x_3$, \dots, $x_{8k+1}$. Under these
assumptions, we have
\begin{equation}
   \Dslash_{8k+2}^\dagger\Dslash_{8k+2}
   =\Dslash_2^\dagger\Dslash_2\otimes1
   +1\otimes\Dslash_{8k}^\dagger\Dslash_{8k},
\end{equation}
and the equation $\Dslash_{8k+2}\Psi=0$ can be solved through the ``separation
of variables'' $\Psi=\xi(x_0,x_1)\otimes\phi(x_2,\ldots,x_{8k+1})$, where
$\Dslash_2\xi=0$ and $\Dslash_{8k}\phi=0$. From Eq.~(\ref{twentytwo}), a
right-handed zero mode $\Gamma^{(8k+2)}\Psi=+\Psi$ is obtained if
(i)~$\sigma^3\xi=+\xi$ and~$\gamma^{(8k)}\phi=+\phi$, or
(ii)~$\sigma^3\xi=-\xi$ and~$\gamma^{(8k)}\phi=-\phi$. Let us denote the
number of $\xi$ satisfying $\sigma^3\xi=+\xi$ by $n$. As noted above, $n$ is
odd and the same as the number of $\xi$ satisfying $\sigma^3\xi=-\xi$. Let us
further denote the number of~$\phi$ satisfying $\gamma^{(8k)}\phi=\pm\phi$ by
$N_\pm$; i.e., $N_+-N_-=\ind\{i\Dslash_{8k}\}$. From these facts, we see that
the number of right-handed zero modes of $\Dslash_{8k+2}$ is given by
\begin{equation}
   n(N_++N_-)=n(2N_-+\ind\{i\Dslash_{8k}\}).
\label{twentyfour}
\end{equation}
Since $n$ is odd, the number of right-handed zero modes of $\Dslash_{8k+2}$
is odd if $\ind\{i\Dslash_{8k}\}$ is odd.

We are now led to ask if there is a gauge-field configuration which provides an
odd index for $\Dslash_{8k}$. In fact there is, and such a configuration can be
constructed by using the so-called $\SO(8)$ instanton in
8~dimensions~\cite{Grossman:1984pi}. The gauge potential 1-form of the
$\SO(8)$ instanton on $S^8$ is given, after the stereographic projection
to $\mathbb{R}^8$, by
\begin{equation}
   A(y)={y^2\over y^2+\rho^2}\,h^{-1}(y)\rmd h(y),
   \quad y^2\equiv\sum_{\alpha=1}^8y_\alpha y_\alpha,
\label{twentysix}
\end{equation}
where $y_\alpha$ denote the orthogonal coordinate system of $\mathbb{R}^8$
and $\rho$ is the size of instanton. The instanton is asymptotically
pure-gauge with the gauge transformation
\begin{equation}
   h(y)={i\sum_{m=1}^6y_m\sigma_m+iy_7\sigma+y_8\over(y^2)^{1/2}}
   \in\SO(8),
\end{equation}
where $\sigma_m$ represents the gamma matrices in 6~dimensions, which satisfy
$\{\sigma_m,\sigma_n\}=2\delta_{mn}$
($m$ and $n$ take the values 1, 2, \dots, 6)
and~$\sigma=-i\sigma_1\sigma_2\cdots\sigma_6$. The gamma matrices are
$8\times8$ and taken to be purely imaginary. From the gauge potential, we have
the field strength 2-form $F=\rmd A+A\wedge A$,
\begin{equation}
   F(y)={2\rho^2\over(y^2+\rho^2)^2}\,\sum_{\alpha,\beta=1}^8
   \sigma_{\alpha\beta}\,\rmd y_\alpha\wedge\rmd y_\beta,
\end{equation}
where the $8\times8$ real and anti-symmetric matrices $\sigma_{\alpha\beta}$
are
\begin{eqnarray}
   &&\sigma_{mn}={1\over4}[\sigma_m,\sigma_n],\quad
   \sigma_{m7}={1\over2}\sigma_m\sigma=-\sigma_{7m},
\nonumber\\
   &&\sigma_{m8}=-{i\over2}\sigma_m=-\sigma_{8m},\quad
   \sigma_{78}=-{i\over2}\sigma=-\sigma_{87}.
\end{eqnarray}
They form the $\SO(8)$ Lie algebra in the $\mathbf{8_s}$ representation. The
field strength is self-dual in the sense that $*(F\wedge F)=+F\wedge F$ and
the instanton has a unit 4th Chern number
\begin{equation}
   {1\over(2\pi)^44!}\int_{S^8}\tr\{F^4\}=+1.
\label{thirty}
\end{equation}

Using this $\SO(8)$ instanton on~$S^8$, we construct the gauge-field
configuration in $8k+2$ dimensions as follows. We take our base manifold as
$S^1\times S^1\times\prod_{i=1}^k\{S^8\}$ and denote the coordinates of the
first $S^1\times S^1$ by~$(x_0,x_1)$ and the (stereographically projected)
coordinates of the $i$th $S^8$ collectively by $y^{(i)}$. On the $i$th
$S^8$, we place the $\SO(8)$ instantons~$A(y^{(i)})$ that take values in
the $i$th factor of the gauge group~$\SO(8)^k$. From the standard index
theorem,
\begin{equation}
   \ind\{i\Dslash_{8k}\}={1\over(2\pi)^{4k}(4k)!}\int\tr\{F^{4k}\},
\label{twentyfive}
\end{equation}
which is valid for our base manifold~$\prod_{i=1}^k\{S^8\}$, and
Eq.~(\ref{thirty}), we see that that the above configuration gives
$\ind\{i\Dslash_{8k}\}=+1$. Finally, from Eq.~(\ref{twentyfour}), we see that
the number of right-handed zero modes of $\Dslash_{8k+2}$ is odd with the
above gauge-field configuration.

Returning to our original problem, the above configuration for $x_0\to+\infty$
corresponds to $A^g$ in~Eq.~(\ref{seven}) and can be written as
$g(x_1)^{-1}(\rmd+\widetilde A)g(x_1)$ with a certain gauge
potential~$\widetilde A$. Combining all the above arguments, we infer that
the Majorana Pfaffian~(\ref{three}) changes sign under the gauge
transformation~(\ref{fourteen}) in the presence of $k$~$\SO(8)$ instantons
along $8k$~dimensions. This phenomenon can properly be regarded as the gauge
anomaly, because, as in the $k=0$ case, this anomaly can be removed only by
sacrificing the perturbative gauge invariance.

In two aspects, the gauge anomaly observed above in the case of higher
dimensions is different from the Witten global gauge anomaly. First, we have
shown the non-invariance
$\Pf\{i\Dslash_{8k+1}\}[A^g]=-\Pf\{i\Dslash_{8k+1}\}[A]$ for the gauge
transformation $g(x_1)$ in Eq.~(\ref{fourteen}). However, the gauge
transformation~$g(x_1)$ does not approach the identity even at the
infinity of~$\mathbb{R}^{8k}$ (in the sense of the stereographic projection),
and the transformation is not localized in all directions. In conventional
treatments of the global gauge anomaly, gauge transformations are restricted
to a class of transformations which approach the identity at the spacetime
infinity. Such gauge transformations in $d$-dimensional Euclidean space
are classified by $\pi_d(G)$, where $G$ is the gauge group.
Second, the gauge field with which we have shown the non-invariance requires
the presence of instantons that cannot be smoothly deformed to $A=0$. This
situation is also different from that for the Witten global gauge anomaly,
which exists even for $A=0$. The $\SO(8)$ instanton given in
Eq.~(\ref{twentysix}) neither is a solution of the Euclidean Yang-Mills theory
$\int\tr\{F\wedge*F\}$ nor has a finite action in~$\mathbb{R}^8$. There is a
possibility, however, that such a configuration becomes relevant in a theory
with higher-dimensional operators in the action, possibly in the context of
string theory. (See, for example, Ref.~\citen{Minasian:2001ib}.)

\section{Conclusion}
In this paper, we have demonstrated a sort of global gauge anomaly
associated with the Majorana fermion in $8k+1$~dimensions. Possible
implications of this phenomenon should be investigated. We expect
interesting applications in quantum mechanics ($k=0$) and in string theory
($k=1$).

\section*{Acknowledgements}
We would like to thank Hiroto So for discussions at the early stage of this
work and Kazutoshi Ohta, Tadakatsu Sakai and Kenichi Shizuya for helpful
discussions. This work is supported in part by MEXT Grants-in-Aid for
Scientific Research (Nos.~13135203, 13135223 and~15740173).

\end{document}